\documentclass[
  ,draft            
  ]
  {aipproc}

\layoutstyle{8x11double}

\usepackage{color}
\usepackage{mdframed}

\definecolor{red}{rgb}{1,0,0}
\definecolor{orange}{rgb}{1,0.5,0}
\definecolor{green}{rgb}{0.13,0.55,0.13}
\definecolor{purple}{rgb}{0.5,0,1}
\definecolor{mag}{rgb}{1.0,0.0,1.}

\begin{document}

\title{Assessing Student Learning in Middle-Division Classical Mechanics/Math Methods}


\author{Marcos D. Caballero}{
  address={Department of Physics, University of Colorado, Boulder, CO 80309}
  }

\author{Steven J. Pollock}{
  address={Department of Physics, University of Colorado, Boulder, CO 80309}
  }

\begin{abstract}
Reliable and validated assessments of introductory physics have been instrumental in driving curricular and pedagogical reforms that lead to improved student learning. As part of an effort to systematically improve our sophomore-level Classical Mechanics and Math Methods course (CM 1) at CU Boulder, we are developing a tool to assess student learning of CM 1 concepts in the upper-division. The Colorado Classical Mechanics/Math Methods Instrument (CCMI) builds on faculty-consensus learning goals and systematic observations of student difficulties. The result is a 9-question open-ended post-test that probes student learning in the first half of a two-semester classical mechanics / math methods sequence. In this paper, we describe the design and development of this instrument, its validation, and measurements made in classes at CU Boulder.
\end{abstract}

\maketitle

\section{\label{sec:intro}Introduction}

Physics educators have investigated student learning in introductory physics over the last three decades \cite{McDermott:1999tz,Meltzer:2012eg}. These investigations were driven in part by data collected from research-based assessment instruments such as the Force Concept Inventory \cite{1992PhTea..30..141H}. Such assessments have been instrumental in helping to identify common student difficulties. Furthermore, results from these instruments have supported curricular and pedagogical transformations in introductory physics and have provided evidence of the success of these transformations. We lack such research-based assessments for our upper-level classical mechanics courses. Research into student learning in middle- and upper-division physics has begun \cite{Meltzer:2012eg, ambrose2004investigating, christensen2009student, Caballero:2012wr,2012arXiv1207.1283W, pepper1289our,Singh:2006tv,Smith:2010wx}, but is far less mature than similar research in introductory physics \cite{McDermott:1999tz, Meltzer:2012eg}. 

At CU Boulder (CU), we are transforming the first half of our two-semester classical mechanics sequence (CM 1), including developing consensus learning goals, investigating student learning, and creating student-centric instructional materials \cite{Pollock:2012uy,CMweb}. In order to assess the transformed course and to help further investigate student difficulties at this level, we have begun to develop an instrument that probes student learning. Here, we present the development of the Colorado Classical Mechanics/Math Methods Instrument (CCMI) and initial investigations into its validity and reliability.

\vspace*{-14pt}
\section{\label{sec:ccmi}The CCMI}

\begin{figure*}

\centering

\begin{minipage}{2.1\linewidth}
\begin{mdframed}
{\bf Learning goals, {\it students should be able to}}:\\
$\cdot$ choose appropriate area and volume elements to integrate over a given shape.\\
$\cdot$ translate the physical situation into an appropriate integral to calculate the gravitational force at a particular point away from some simple mass distribution.
\end{mdframed}
\vspace*{5pt}
\begin{minipage}{0.72\linewidth}
{\bf Q9} Consider an infinitely thin cylindrical shell with non-uniform mass per unit area of $\sigma (\phi, z)$. The shell has height $h$ and radius $a$, and is not enclosed at the top or bottom.

\medskip

(a) What is the area, $dA$, of the small dark gray patch of the shell which has height $dz$ and subtends an angle $d\phi$ as shown to the right?\\
(b) Write down (BUT DO NOT EVALUATE) an integral that would give you the MASS of the entire shell. Include the limits of integration.
\end{minipage}
\begin{minipage}{0.25\linewidth}
\flushright
\includegraphics[clip,trim=70mm 40mm 70mm 30mm,width=0.75\linewidth]{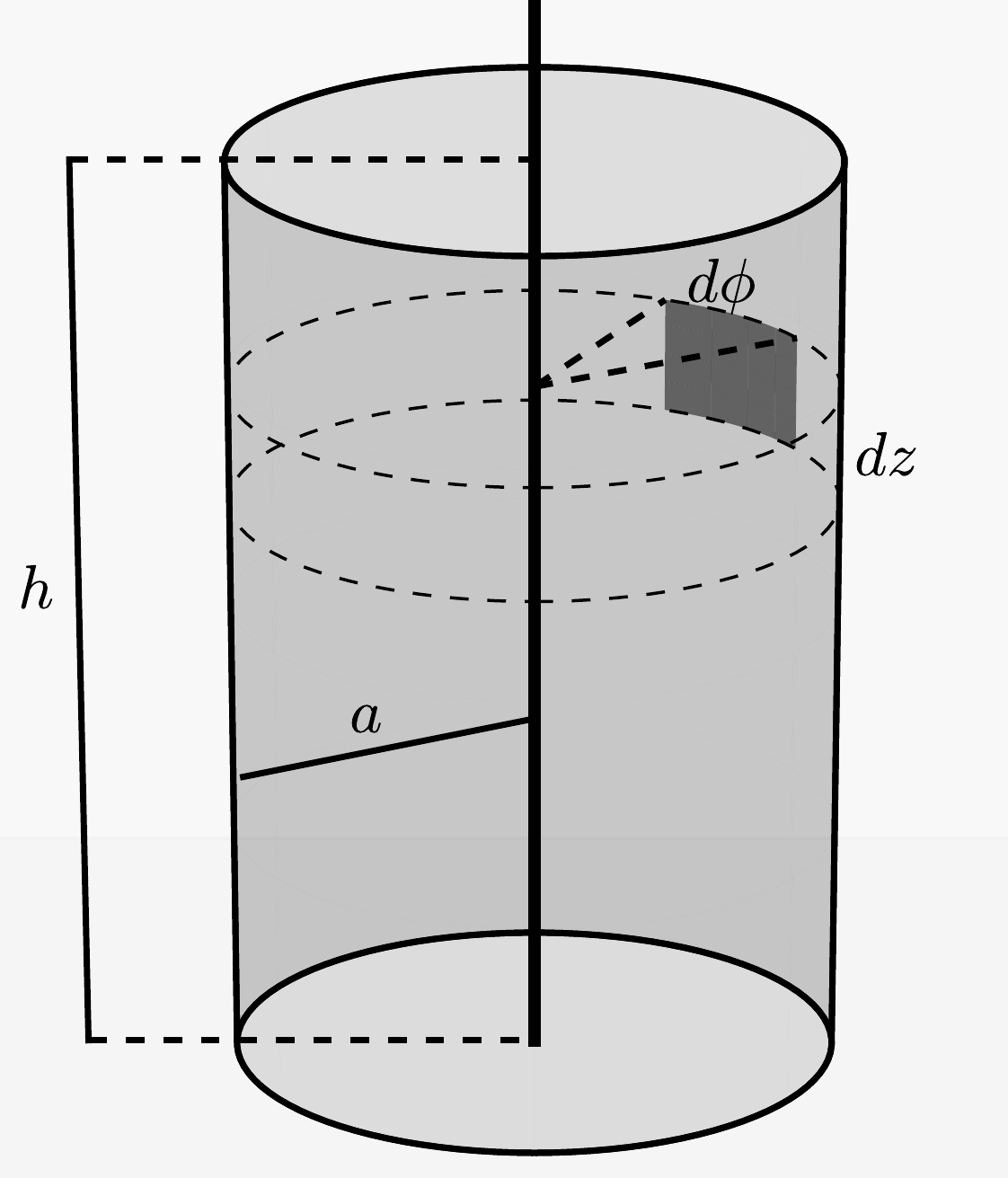}
\end{minipage}
\end{minipage}

\caption{Certain topic-scale learning goals are evaluated by the CCMI questions. The sample question appears on the CCMI pre- and post-tests; vector calculus is a prerequisite for CM 1. This question constitutes 9\% of the total post-test score.}\label{fig:mass}

\end{figure*}

The CCMI is a 9-question open-ended test that focuses on topics taught in the first half of a two-semester classical mechanics sequence. This first course concludes before a discussion of the calculus of variations; hence, the Lagrangian and Hamiltonian formulations of mechanics are absent from the test. The CCMI focuses on core skills and commonly encountered problems. Students solve a variety of problems such as: determining the general solution to common differential equations (e.g., $\ddot{x}=-A^2x$); finding equilibria and sketching net forces on a potential energy contour map; and decomposing vectors in Cartesian and plane-polar coordinates. We have designed the CCMI to be given in a standard 50-minute lecture period. To accompany the longer post-test, we have developed a short (15-20 minute) pre-test that contains a subset of three problems gleaned from the post-test. Figure \ref{fig:mass} shows a sample CCMI question that appears on both the pre- and the post-test. 

\vspace*{-14pt}
\section{\label{sec:dev}Instrument Development}

{\bf Writing Questions}: As the initial step towards transforming CM 1, a series of faculty meetings were held to develop consensus course-scale learning goals and to articulate the topical content coverage of the course \cite{pepper2012facilitating}. After the development of course-scale learning goals, a set of specific, topical learning goals were drafted. To develop these learning goals, we utilized field notes collected during lectures, weekly homework help sessions, and faculty meetings. A further set of faculty meetings were held in which the topical learning goals were agreed upon. In these meetings, several topical learning goals were selected to be assessed on the CCMI. These course-scale and topical-scale learning goals are available online \cite{CMweb}. Based on these topical learning goals determined by the faculty, sixteen open-ended questions were initially written.  Some of these questions were adapted from exam or clicker questions written by CU faculty in previous semesters. All questions were informed by observed student difficulties \cite{Pollock:2012uy}. 

{\bf Expert Validation}: Initially, two CU faculty members who had recently taught the course reviewed the sixteen CCMI questions for clarity and content. In working to establish the validity of the CCMI, these faculty aimed to answer two questions: (1) Does each question address the concepts and skills that my students should master? (2) Is each question written in a clear and concise manner? Over the course of the next two years, faculty meetings were held to discuss CCMI questions and students' responses to each question were scrutinized. Additional individual feedback was solicited from faculty at CU and elsewhere. Questions were iteratively improved between each administration of the instrument, and some questions were cut altogether. Over twenty faculty members provided input leading to a 9-question version of the CCMI that most students are able to complete in a 50-minute period. The latest version of the CCMI (v3.0) is available online \cite{CMweb}.

{\bf Student Validation}: In parallel to incorporating feedback from physics faculty, several rounds of student interviews were conducted to ensure the instrument probed persistent challenges and to determine if students were interpreting questions as we intended. After solving the full set of problems on the CCMI in a think-aloud setting, interviewees were asked to discuss their reasoning for specific answers in more detail. This process provided additional information about which questions should remain and which should be removed from each version of the CCMI. In addition, student interviews were critical to establishing concise wording and a clear focus for each question. Muddled responses or clarifying questions from interviewees were strong indicators that the wording or focus of a particular question needed to be reconsidered. After each round of interviews, an updated version of the CCMI was constructed. In total, fourteen physics students at CU were interviewed while solving the CCMI.

\vspace*{-14pt}
\section{\label{sec:grading}Grading Rubric Construction}

Once questions were finalized, we developed a detailed grading rubric for the CCMI that was informed by student work. For example, common student responses to question Q9(a) (Figure \ref{fig:mass}) included: (1) $a\,d\phi\,dz$ , (2) $r\,d\phi\,dz$, (3) $d\phi\,dz$, and (4) $r\,dr\,d\phi\,dz$. Full credit is awarded to the first response. Partial credit is earned for the second response. Here, students neglected to use the length-scale given in the problem statement and substituted the radial coordinate. However, interviews suggest that students treated $r$ as a constant, and were likely to construct an appropriate two-dimensional integral. No credit is awarded for the final two responses. The third has incorrect units and the last is a volume (not area) element. In this way, the rubric emphasizes concept mastery; partial credit is only awarded for minor mistakes.

For other questions, student responses are more varied; there were over 40 unique responses to the latest version of Q9(b) (Figure \ref{fig:mass}). The development of a common rubric for such questions can quickly become overly complex. For example, the grading rubric developed for the Colorado Upper-division Electrostatics diagnostic (CUE) \cite{Chasteen:2012fl} requires formal training. By emphasizing concept mastery, we avoided constructing a complicated grading rubric. For this question, the solution was decomposed into its constituent parts (i.e., a double integral over $d\phi$ and $dz$, correct limits on each integral, the inclusion of the mass density, an appropriate kernel, etc.). We were then able to develop a simple rubric in which constituents were graded.

With a mastery-focused rubric, we aim to decouple two important purposes for these types of assessments: (1) evaluating performance and (2) gaining insight into student learning. We are developing a complementary coding scheme that captures student difficulties on each question. Coding student work on the CCMI has helped identify common and persistent difficulties in key areas of classical mechanics \cite{Caballero:2012wr}. By separating the two roles, educators and researchers can choose the lens through which they want to view students' responses to CCMI questions based on their own interests and time. Results from this coding scheme will be the subject of a future publication.

\vspace*{-14pt}
\section{\label{sec:ruler}Results \& Test Statistics}

The most recent version of the CCMI was administered at CU for the last three semesters (N=167). One PER faculty member (SJP) and two traditional research faculty taught CM 1 using a variety of pedagogical techniques, which were developed as part of the larger course transformation project \cite{Pollock:2012uy,CMweb}. Table \ref{tab:cu} briefly summarizes these pedagogies along with the number of students who took the CCMI post-test and the mean score earned in each class.

\begin{table}
\begin{tabular}{cllcc}\hline\hline
{\bf Sem.} & {\bf Faculty} & {\bf Pedagogy} & {\bf N} & {\bf CCMI (\%)} \\\hline
1 & PER & CQ, T, GP, L & 62 & 59.7 $\pm$ 2.8 \\
2 & TRAD & CQ, T, GP, L & 41 & 46.1 $\pm$ 3.0 \\
3 & TRAD & CQ, L & 67 & 51.0 $\pm$ 2.5 \\\hline
\end{tabular}\label{tab:cu}
\caption{Courses taught at CU all involved different forms of engagement including Clickers (CQ), in-class Tutorials (T), Group Problem sessions (GP), and Lecture (L).}
\end{table}

\begin{figure}
\includegraphics[width=0.95\linewidth, clip, trim=0mm 0mm 10mm 10mm]{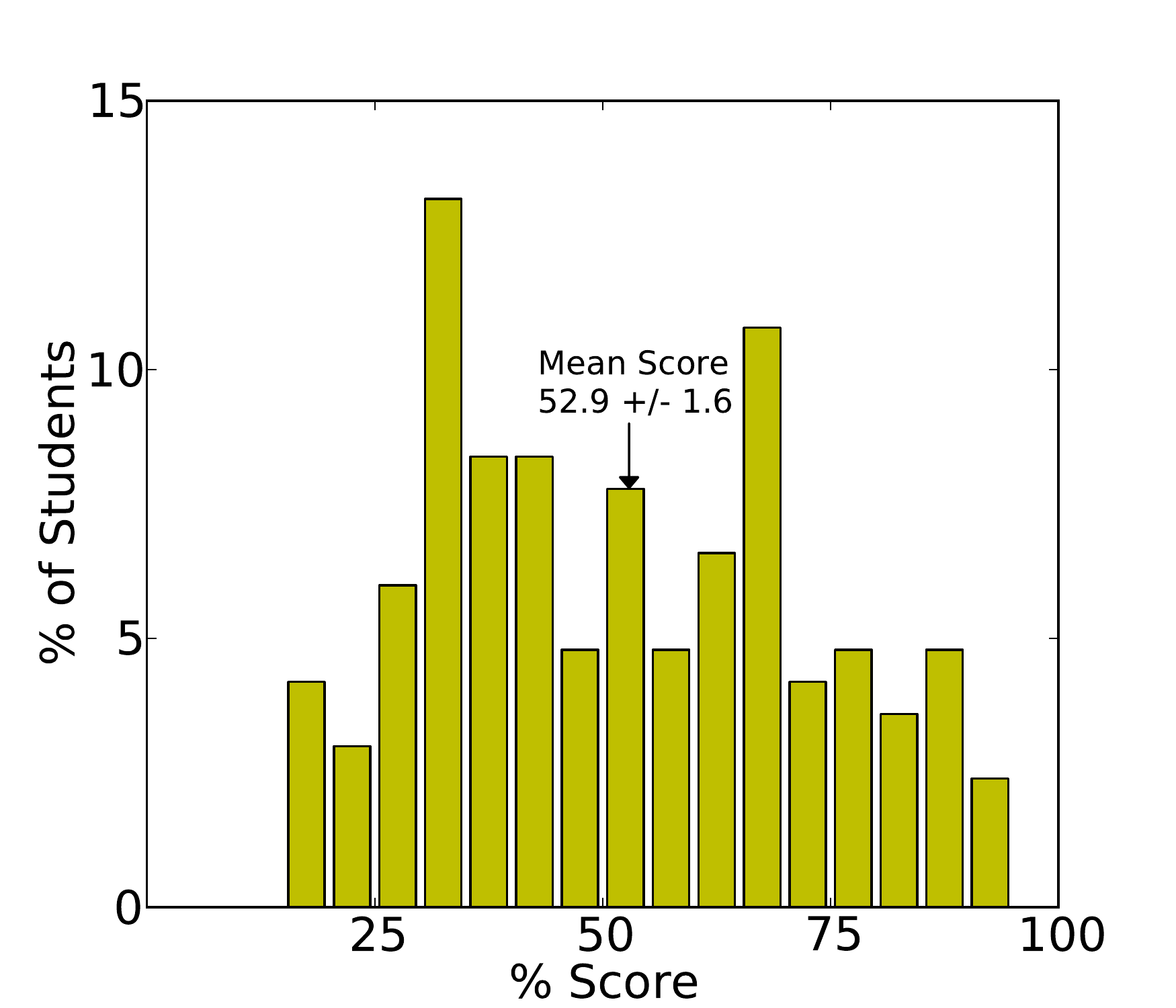}
\caption{Distribution of CCMI post-test scores (N=167).}\label{fig:dist}
\end{figure}

Ultimately, we aim to develop an instrument that can help evaluate instructional strategies and curricular transformations like those presented above. To that end, we must develop a valid, reliable, and internally consistent instrument. Score distributions for each class were normal (or nearly so), the variances of each distribution were similar, and all course instructors made use of some transformed materials, which justify pooling the data to consider these issues. For the pooled data, the mean score over all three semesters was 52.9 $\pm$ 1.6 \%. The CCMI is a challenging test, and the grading rubric is strict. However, CM 1 students earned a wide range of scores (Fig.\ \ref{fig:dist}); some outperformed a sample of first-year graduate students at CU (avg. 74.5 $\pm$ 3.4 \%, N=5).

Looking at the three courses individually, scores in semesters 1 and 2 were normally distributed, while those in semester 3 were slightly non-normal ($A^2=0.76$, $p<0.05$) \cite{anderson1954test}. A Kruskall-Wallis test detected a group difference ($H=10.3$, $p<0.05$) \cite{kruskal1952use}, and a series of pairwise Mann-Whitney tests \cite{mann1947test} with family-wise error control ($\alpha = 0.017$) \cite{dunnett1955multiple} demonstrated that students taught by the PER faculty member (who was actively developing CM 1 course materials) outperformed students in both other courses. Additional testing in transformed and non-transformed courses at CU and elsewhere is needed to form clear conclusions about the effect of pedagogy and instructors on CCMI post-test scores. For example, in semester 2, which is the off-semester for CM 1, we found that students earned pre-test scores that were 9.5 and 6.7 points lower than students earned during on-semesters 1 and 3, respectively.

{\bf Criterion Validity}: We have established, at least at CU, the face and content validity of the CCMI (Sec.\ \ref{sec:dev}), but establishing the instrument's criterion validity is equally important. Students' exams are the most similar measure to the CCMI. Like exams, the CCMI is completed individually in timed and controlled environments. But, unlike exams, it does not affect students' grades. Each class took three exams: two regular hour exams and a final. The averages of those three exams were z-scored to allow comparisons of different instructors. CCMI post-test scores were strongly correlated with these z-scored exam averages ($r=0.71$, $p<0.05$); a linear model can thus account for $50$\% of the variance in exam scores associated with CCMI scores. Similarly high correlations were observed on the CUE \cite{Chasteen:2012fl}.

\begin{figure}
\includegraphics[width=0.95\linewidth, clip, trim=0mm 0mm 10mm 10mm]{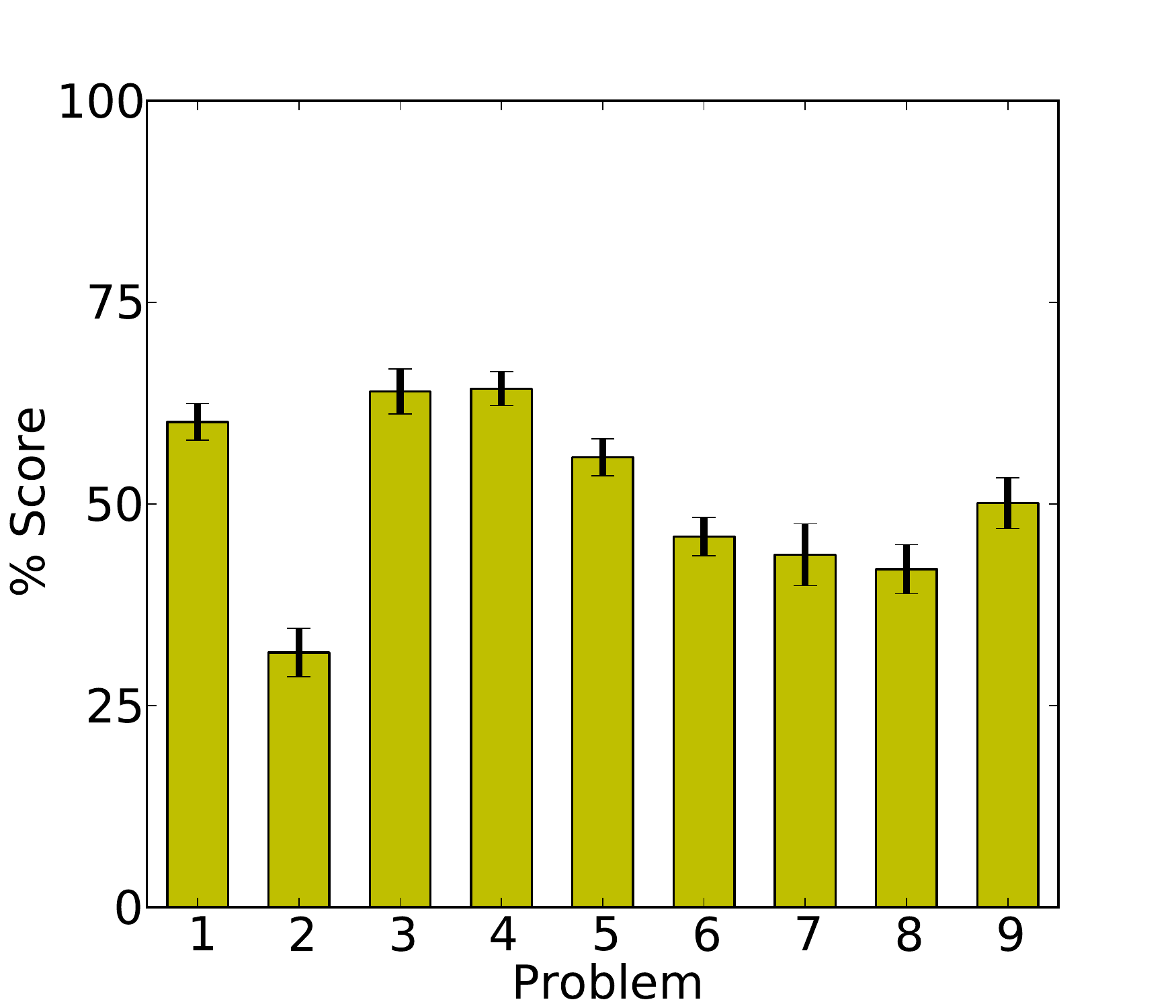}
\caption{Performance on each CCMI item (N=167).}\label{fig:item}
\end{figure}

{\bf Item-test Correlation}: In addition to overall CCMI scores connecting well to external measures, it is desirable for individual items to connect well to the rest of the test. Individual items on the CCMI pose different challenges to students, and performance varies from 30\% to 65\% (Fig.\ \ref{fig:item}). Even so, we find that performance on individual items generally correlates well with the test overall. The item-test correlation for each question varies between 0.45 and 0.53 with exception of question 2 ($r=0.31$). While there is no widely accepted cutoff for item-test correlation, a common criteria is $r\geq0.2$ \cite{Ding:2006kq}, which all CCMI items achieve. The low correlation of question 2 is likely due to generally lower performance on this item. Question 2 covers Taylor series, a challenging topic for our sophomore students \cite{Caballero:2012wr}.

{\bf Internal Consistency}: Items must also give consistent results with one another. Cronbach's alpha measures the degree to which test items measure related constructs, that is, the degree of internal consistency of the test. Cronbach's alpha for the CCMI is 0.77, which is just below the low-stakes testing cutoff of 0.80. Cronbach's alpha depends strongly on the sample used to compute it \cite{wallace2010concept} and, thus, future data will provide a better estimate of the true alpha. However, the CCMI does not measure a single construct, which violates the underlying assumptions of Cronbach's alpha. Thus, our computed Cronbach's alpha underestimates the true alpha and therefore internal consistency of CCMI is still high \cite{graham2006congeneric}.

\vspace*{-14pt}
\section{\label{sec:closing}Concluding remarks}

While the CCMI is still under development and refinement, preliminary results suggest it is a valid and reliable instrument for investigating student learning in middle-division classical mechanics / math methods at CU. The mastery-focused grading philosophy provides a relatively simple rubric, which requires no formal training. The complementary coding scheme helps separate the dual roles of the assessment; it is both a tool for evaluation and a window into student thinking. At CU, student performance on individual items on the CCMI has helped set the research agenda for investigations into student thinking around Taylor series.

While presenting the results from the CCMI, we avoided a discussion of the data collected from eight partner institutions. This is first because most partner institutions have low enrollment in their classical mechanics courses; between 3 and 14 students took the post-test at these schools. Secondly, CM 1 is a combined classical mechanics / math methods course, which is an uncommon combination. Most of these partner institutions offer a single semester course that covers through Hamiltonian dynamics or do not explicitly teach mathematical methods in the first half of their two semester sequence. The CCMI represents the learning goals emphasized in our combined classical mechanics / math methods course. Therefore detailed discussions with partner faculty are needed to help frame the results from their institutions. Future research will address these concerns as we work to assist these instructors in evaluating their classical mechanics courses.

\vspace*{-14pt}
\begin{theacknowledgments}
We gratefully acknowledge the generous contributions of CU faculty, especially A.D. Marino, J.L. Bohn, K.P. McElroy, and collaborating faculty elsewhere. Particular thanks to the members of PER@C, including former member R.E. Pepper who designed the original CCMI. We also greatly appreciate the help of our student participants. This work was supported by University of Colorado's Science Education Initiative.
\end{theacknowledgments}



\bibliographystyle{aipproc}   

\bibliography{ccmi}

\IfFileExists{\jobname.bbl}{}
 {\typeout{}
  \typeout{******************************************}
  \typeout{** Please run "bibtex \jobname" to optain}
  \typeout{** the bibliography and then re-run LaTeX}
  \typeout{** twice to fix the references!}
  \typeout{******************************************}
  \typeout{}
 }

\end{document}